\documentclass[letter]{aa} 
\usepackage{graphicx}
\usepackage{txfonts}

\usepackage{natbib}
\bibpunct{(}{)}{;}{a}{}{,}
\begin{document}
   \title{Evidence of enrichment by individual SN from elemental
     abundance ratios in the very metal-poor dSph galaxy 
     Bo\"otes\,I
\thanks{The data presented herein were obtained at the
       W.M. Keck Observatory, which is operated as a scientific
       partnership among the California Institute of Technology, the
       University of California and the National Aeronautics and Space
       Administration. The Observatory was made possible by the
       generous financial support of the W.M. Keck Foundation.}}

   \subtitle{}
\titlerunning{Abundances in the Bo\"otes dSph galaxy}
   \author{S. Feltzing
          \inst{1}
          \and
          K. Eriksson \inst{2} 
          \and
         J. Kleyna \inst{3}
          \and
          M.I. Wilkinson \inst{4}
          }

\authorrunning{S. Feltzing et al.}

   \offprints{S. Feltzing}

   \institute{Lund Observatory, Box 43, SE-221 00 Lund, Sweden
              \email{sofia@astro.lu.se}
         \and
Department of Astronomy and Space Physics,   Uppsala University,   Box 515, 
  SE-751 20 Uppsala,  Sweden \email{kjell.eriksson@astro.uu.se} \and
Institute for Astronomy, Honululu, 2680 Woodlawn Drive, Honolulu, HI 96822, USA  \and
Department of Physics and Astronomy, University of Leicester,
  University Road, Leicester LE1 7RH, UK \email{miw6@astro.le.ac.uk}
             }

   \date{Accepted for publication in A\&A Letters}

 
  \abstract
   {}
{We establish the mean metallicity from
   high-resolution spectroscopy for the recently found dwarf
   spheroidal galaxy Bo\"otes\,I and test whether it is a common feature
   for ultra-faint dwarf spheroidal galaxies to show signs of
   inhomogeneous chemical evolution (e.g. as found in the Hercules
   dwarf spheroidal galaxy).}
{We analyse high-resolution, moderate
   signal-to-noise spectra for seven red giant stars in the Bo\"otes\,I
   dSph galaxy using standard abundance analysis techniques. In
   particular, we assume local thermodynamic equilibrium and employ
   spherical model atmospheres and codes that take the sphericity of
   the star into account when calculating the elemental abundances.}
{We confirm previous determinations of
   the mean metallicity of the Bo\"otes\,I dwarf spheroidal galaxy to be --2.3 dex.
   Whilst five stars are clustered around this metallicity, one is
   significantly more metal-poor, at --2.9 dex, and one is more
   metal-rich at, --1.9 dex. Additionally, we find that one of the stars,
Boo-127, shows an atypically high [Mg/Ca] ratio, indicative
of  stochastic enrichment processes within the dSph galaxy. Similar results
have previously only been found in the Hercules and Draco dSph galaxies
and appear, so far, to be unique to this type of  galaxy.
}
   {}


   \maketitle
%
\section{Introduction}

Until recently, the number of dwarf spheroidal (dSph) galaxies around
the Milky Way was small compared to expectations from $\Lambda$CDM
\citep{moore1999}. However, in the past few years several new systems
have been found through systematic searches
\citep[e.g.][]{belokurov.boo,belokurovcats}.

In general, dSph galaxies are some of the most tenuous stellar systems
that we know of.  This is especially true for the newly found dSph
galaxies which have very low stellar luminosities \citep[see
e.g., ][]{martin08}. The new dSphs are ultra-faint and show low
metallicities as indicated by low-resolution spectroscopy
\citep{kirbyletter,kochbierman}. \citet{koch2008Her} found unusual
abundance patterns in two red giant stars (RGB) in the ultra-faint
Hercules dSph galaxy. Because of the low baryonic mass for these
system it has been speculated that the elemental abundances in the
stars in these systems might show us the results of individual
supernova events \citep{koch2008Her}.

The recently found dSph galaxy in Bo\"otes \citep[Bo\"otes\,I,
][]{belokurov.boo} provides an excellent opportunity to test whether
or not unusual elemental abundance ratios are a common feature of
ultra-faint dSph galaxies, thanks to its low baryonic mass,
\citet{belokurov.boo} estimate, based on a colour magnitude diagram,
that the Bo\"otes\,I dSph galaxy is a purely old and metal-poor
system.  Low-resolution spectroscopic data confirm this
\citep{martin07,norris2008} at find $<$[Fe/H]$>$=--2.5.

With $M_{\rm V} \sim -5.8$ this dSph galaxy is one of the least
luminous galaxies known \citep{belokurov.boo}. \citet{fellhauer2008},
modelled the system and find that if this galaxy ever had a dark
matter halo, it must still have it.  This implies that, since the dark
matter provides a deep potential well, the stars that originally
formed in the dSph galaxy are still there and, moreover, the depth of
the well should have helped retain the ejecta from core-collapse
supernova. For Hercules, \citet{koch2008Her} conclude that about 10
supernova are needed to pollute the interstellar medium to the
observed atypical abundance ratios. Given that Bo\"otes\,I has an even
lower baryonic mass than Hercules, we might expect to be able to see
enrichment from individual supernovae in the elemental abundance
trends (which would show up as large scatter in element ratios from
star to star).

We have obtained high-resolution, moderate S/N spectra for seven RGB
stars in the Bo\"otes\,I dSph galaxy.  Here we report on the mean
metallicity, the metallicity spread, and atypical abundance ratios
similar to those found in the Hercules \citep{koch2008Her} and Draco
dSph galaxies \citep{fulbright2004Draco}. Thus, Bo\"otes\,I
becomes the third system to show unexpected abundances ratios.

\section{Observations and abundance analysis}
\label{sect:obs}

\begin{figure}
\resizebox{\hsize}{!}{\includegraphics[angle=0]{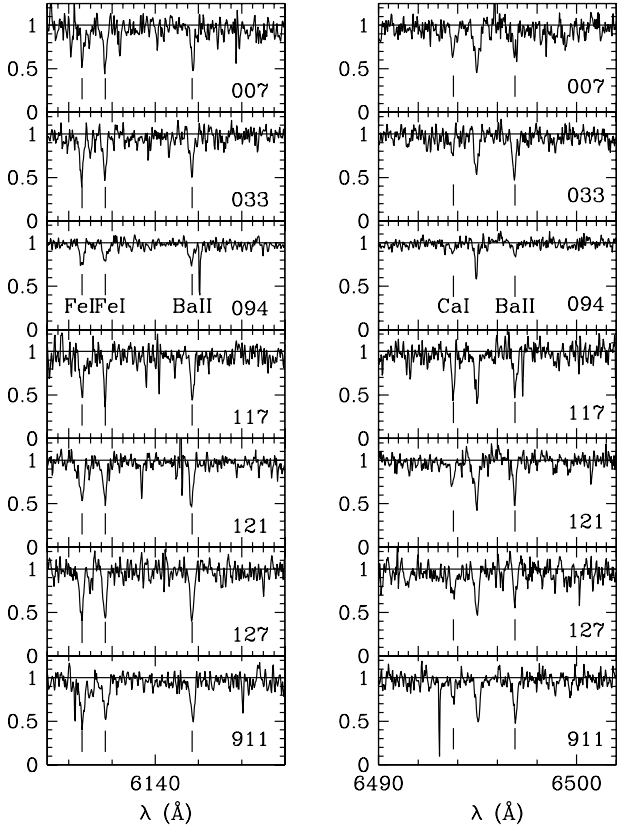}}
\caption{Portions of the stellar spectra around the two Ba\,{\sc ii}
lines used in this study. Those and additional lines measured 
are identified in the panels for Boo-094 (third from the top).
S/N are typically somewhat higher than 20-25. For Boo-094 S/N
reaches 30-35.}
         \label{fig:spectra}
   \end{figure}

\begin{table*}
\caption{Identification of programme stars and the derived elemental abundances.}
\label{stars.tab}      
\begin{tabular}{l l llllllllllllllllll }     
\hline\hline       
ID & R.A. & Dec. &  $V$   &  [Fe/H]  & [Fe/H] & $\epsilon$(Fe) &  $\epsilon$(Mg) & & $\epsilon$(Ca) &  $\epsilon$(Ba)\\ 
   & (J2000.0)     & (J2000.0)     &       & (7.45)  & Norris08 \\\hline 
Boo-007 &  13 59 35 & +14 20 23 &  17.88   & --2.28 & --2.32 & 5.17\,\,0.26\,(22) & 5.68\,\,0.18\,(4) & 5.57 & 4.24\,\,0.22\,(5) & --0.90 (1)\\ 
Boo-033 &  14 00 11 & +14 25 01 &  17.79   & --2.47 & --2.96 & 4.98\,\,0.36\,(25) & 5.53\,\,0.30\,(4) & 5.77 & 4.22\,\,0.28\,(5) & --0.74\,\,0.24\,(2)\\               
Boo-094 &  14 00 31 & +14 34 03 &  16.99   & --2.90 & --2.79 & 4.55\,\,0.22\,(25) & 5.14\,\,0.04\,(4) & 5.12 & 3.61\,\,0.03\,(4) & \\
Boo-117 &  14 00 10 & +14 31 45 &  17.74   & --2.24 & --1.72 & 5.21\,\,0.30\,(21) & 5.66\,\,0.23\,(3) & 5.53 & 4.34\,\,0.28\,(4) & --0.57\,\,0.11\,(2)\\
Boo-121 &  14 00 36 & +14 39 27 &  17.42   & --2.39 & --2.37 & 5.06\,\,0.22\,(26) & 5.53\,\,0.30\,(4) & 5.78 & 4.28\,\,0.29\,(5) & --0.69 (1)\\
Boo-127 &  14 00 14 & +14 35 52 &  17.69   & --1.98 & --1.49 & 5.47\,\,0.34\,(29) & 6.28\,\,0.11\,(3) & 6.33 & 4.33\,\,0.18\,(5) & --0.49\,\,0.43\,(2)\\
Boo-911 &  14 00 01 & +14 36 51 &  17.49   & --2.21 & --1.98 & 5.24\,\,0.29\,(29) & 5.63\,\,0.24\,(4) & 5.45 & 4.48\,\,0.05\,(4) & --0.64\,\,0.09\,(2)\\
\hline                  
\end{tabular}
\end{table*}

Observations using the High Resolution Echelle Spectrometer (HIRES)
\citep{1994SPIE.2198..362V} on Keck\,I were carried out in June
2006. We obtained spectra of reasonable quality for seven RGB stars in
the Bo\"otes\,I dSph galaxy.

\begin{figure}
\resizebox{\hsize}{!}{\includegraphics[angle=0]{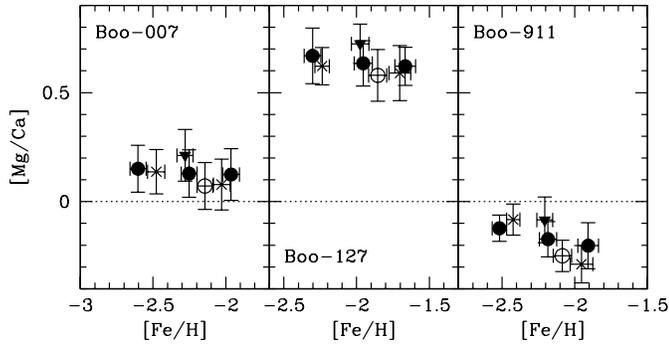}}
\caption{Illustration of the robustness of the measured [Mg/Ca] to
  errors in the stellar parameters. For each star we show the [Mg/Ca]
  ratio using seven model atmospheres all with different stellar
  parameters: $\bullet$ shows three models with different $T_{\rm
    eff}$ (the middle one being the one adopted in the final abundance
  analysis).  A filled, inverted triangle denotes a model with the
  final $T_{\rm eff}$ but with $\log g=0.5$, and $\circ$ denotes a
  model with the finally adopted $T_{\rm eff}$, $\log g =1.0$, and the
  finally adopted $\xi_{\rm t}$ but with [Fe/H]$=-2.50$. Finally,
  $\times$ denotes two models with all parameters set to the finally
  adopted values but with $\xi_{\rm t}$ +0.6 and --0.4\,km\,s$^{-1}$
  relative to the final value. The error-bars denote the error in the
  mean for the derived abundances; e.g., for [Mg/Ca] this is
  equivalent to $\sqrt{\sigma_{\rm Mg}^2/N_{\rm lines, Mg} +
    \sigma_{\rm Ca}^2/N_{\rm lines, Ca}}$.}
\label{params.fig}
\end{figure}

A full description of the construction of the linelist, determination
of stellar parameters, etc. will be given elsewhere. Briefly, we derived
$T_{\rm eff}$ from infrared 2MASS photometry
\citep{2006AJ....131.1163S} using the calibrations by
\citet{houdashelt2000} and \cite{{alonso1999calib}}. We found that all
stars could be modelled using $T_{\rm eff}$= 4600\,K, apart from
Boo-094, which has a somewhat cooler model.  Once $T_{\rm eff}$ was
determined, microturbulence ($\xi_{\rm t}$) was checked by requiring
all Fe\,{\sc i} lines to yield the same Fe abundance regardless of
linestrength. Given the quality of the spectra, a common $\xi_{\rm t}$
of 2.1\,km\,s$^{-1}$ yields a consistent result for all stars. Surface
gravity ($\log g$) was set to 1.0\,dex for all stars, apart from
Boo-094, for which we adopt 0.5\,dex. Boo-094 is clearly more evolved;
e.g., lines sensitive to $\log g$, such as the Ca\,{\sc i} line at
616.2\,nm, show that this star has a low $\log g$. Our results are not
sensitive to the adopted $\log g$ (compare Fig.\ref{params.fig}).

Assuming that local thermodynamical equilibrium will hold, we
performed a standard abundance analysis using MARCS model atmospheres
\citep{gustafsson2008} and accompanying programs for abundance
analysis. These codes take the sphericity of the stellar atmospheres
into account, which is necessary because our stars are quite
evolved. Using nonspherical models results in significant errors
\citep[][]{heitereriksson}.  The calculations of the elemental
abundances incorporate the broadening of the lines through collisions
by neutral hydrogen
\citep{barklem1,barklem2,barklem3,barklem4,barklem5}.

We show two portions of stellar spectra in Fig.\,\ref{fig:spectra}.
Boo-094 clearly has the highest S/N and the weakest spectral
lines. That this star comes out as the most metal-poor in our
abundance analysis is hence not surprising (cf. Table\ref{stars.tab}).

We find it entirely plausible that our [Fe/H] determinations, and
others', can have systematic and/or random errors as large as
$0.3-0.5$\,dex.  We arrive at this conclusion based on results such as
those shown in Fig.\,\ref{params.fig}. In Fig.\,\ref{params.fig} we
investigate how much the abundance ratios change when $T_{\rm eff}$ is
changed by --200\,K and +150\,K, $\log g$ by +0.5 dex, $\xi_{\rm t}$
by +0.6 and --0.4\,km\,s$^{-1}$, and [Fe/H] by --0.5\,dex. Remarkably,
[Mg/Ca] hardly changes, whilst the greatest differences for [Fe/H]
amounts to 0.5\,dex.  However, even given the uncertainty in [Fe/H],
it appears very clear that Boo-094 is significantly more metal-poor
than the bulk of RGB stars in the Bo\"otes\,I dSph galaxy.

From our sample of 7 stars we find a mean iron abundance of 5.1,
on the scale where log N(H) = 12.00, with a
sigma of 0.3 dex. This gives an [Fe/H] of --2.35 dex with the solar
iron abundance taken to be 7.45 \citep{2006NuPhA.777....1A}. This is
compatible with what has previously been found for the system
\citep{norris2008}.  The determinations by \citet{norris2008} are
based on measurements of the Ca H and K lines. The difference in
[Fe/H] is --0.1 dex with a sigma of 0.3 dex (cf.
Table\,\ref{stars.tab}). This must be regarded as excellent
agreement. 

However, a deeper investigation reveals that there is a trend between
our data such that, for low-metallicity stars, our [Fe/H] are higher
than those found in \citet{norris2008}.  It is not clear where this
difference stems from and our data-set is too small to investigate
this further. We note, however, that the difference between us and
\citet{norris2008} also correlates with the equivalent width of the
Ca\,{\sc ii} triplet line at 855.2\,nm as measured by them, such that
the difference is large and positive for smaller equivalent widths
and large and negative for stronger lines. A more extensive
comparison should be made to conclude on the source of the difference
found.

Identifications and elemental abundances for our stars are listed in
Table\,\ref{stars.tab}. The table contains the following in formation:
column one lists the stars' designations as used in this paper and in
\cite{norris2008}. Columns two and three list the right ascension and
declination, respectively. Column four lists the $V$
magnitude. Columns five and six list our [Fe/H] (assuming a solar
[Fe/H] of 7.45) and the [Fe/H] derived by \citet{norris2008}. Columns
seven, eight, ten, and eleven then list our derived abundances as
indicated. For each element, we first list the abundance, the
line-to-line scatter and, then the number of lines in parenthesis. For
Mg (column nine) we give the Mg abundance if the two strong Mg triplet
lines at 5172.6 and 5183.6\,{\AA} are excluded.


\section{The metallicity (spread) in the Bo\"otes\,I dSph galaxy}

It thus appears that the metallicity distribution, as derived from
high-resolution spectroscopy, on the upper RGB of the Bo\"otes\,I dSph
galaxy is dominated by one metallicity with few outliers. That the
outliers are real and not caused by measurement errors is further
demonstrated by inspection of the stellar spectra
(Fig.\,\ref{fig:spectra} and Sect.\,\ref{sect:obs}).  The stars in our
study span about 1\,dex. \citet{norris2008} find a total spread of
about 1.7\,dex and \citet{martin07} a spread of 1.3\,dex. Both of
these studies include more stars than ours.

The position on the sky of our stars do not indicate that, e.g., the
most metal-poor star is at the outskirts of Bo\"otes\,I. In fact, the
overall shape and extent of the Bo\"otes\,I dSph are currently poorly
constrained. The radial velocity selected stars from \citet{martin07}
and \citet{norris2008} appear to have complementary sky coverage. All
our stars are also studied by \citet{norris2008}, so are radial
velocity members (see their Table\,1).

\section{Abundance ratios -- signs of individual supernovae?}
\label{sect:res}

\begin{figure*}
\sidecaption
\includegraphics[width=13cm]{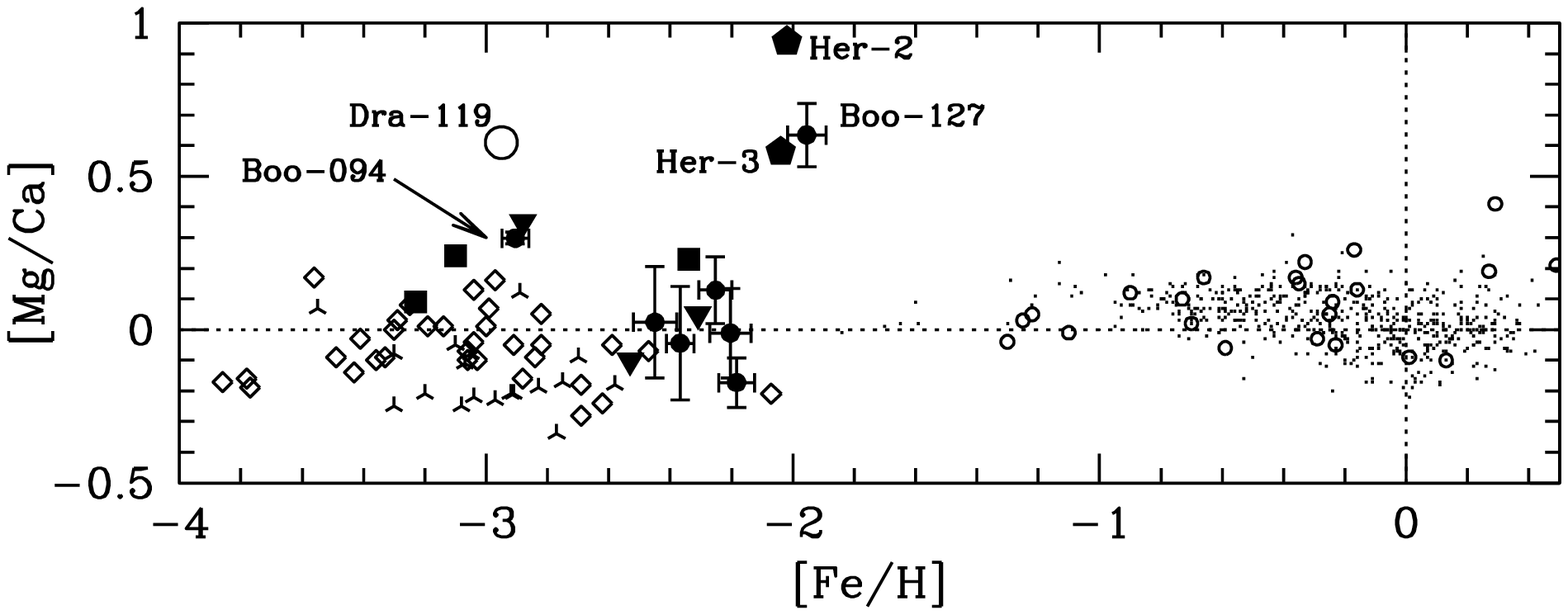}
\caption{[Mg/Ca] as a function of [Fe/H] for Bo\"otes\,I ($\bullet$
  with error-bars), Hercules \citep[solid pentagons,][]{koch2008Her},
  Dra-119 \citep[$\bigcirc$,][]{fulbright2004Draco}, Ursa Major\,II
  (filled squares), and Coma Berenices (filled, inverted triangles)
  \citep[both from][]{frebel2009CBU}. Bulge giants
  \citep[$\circ$,][]{fulbright2007}, Milky Way solar neighbourhood
  dwarf stars ($\cdot$, Bensby et al. in prep.), halo giant stars
  \citep[tripods,][]{cayrel2004}, halo dwarf stars
  \citep[$\diamond$][]{bonifacio2009}. }
     \label{mgca.fig}
\end{figure*}

\begin{figure}
\resizebox{\hsize}{!}{\includegraphics[angle=-90]{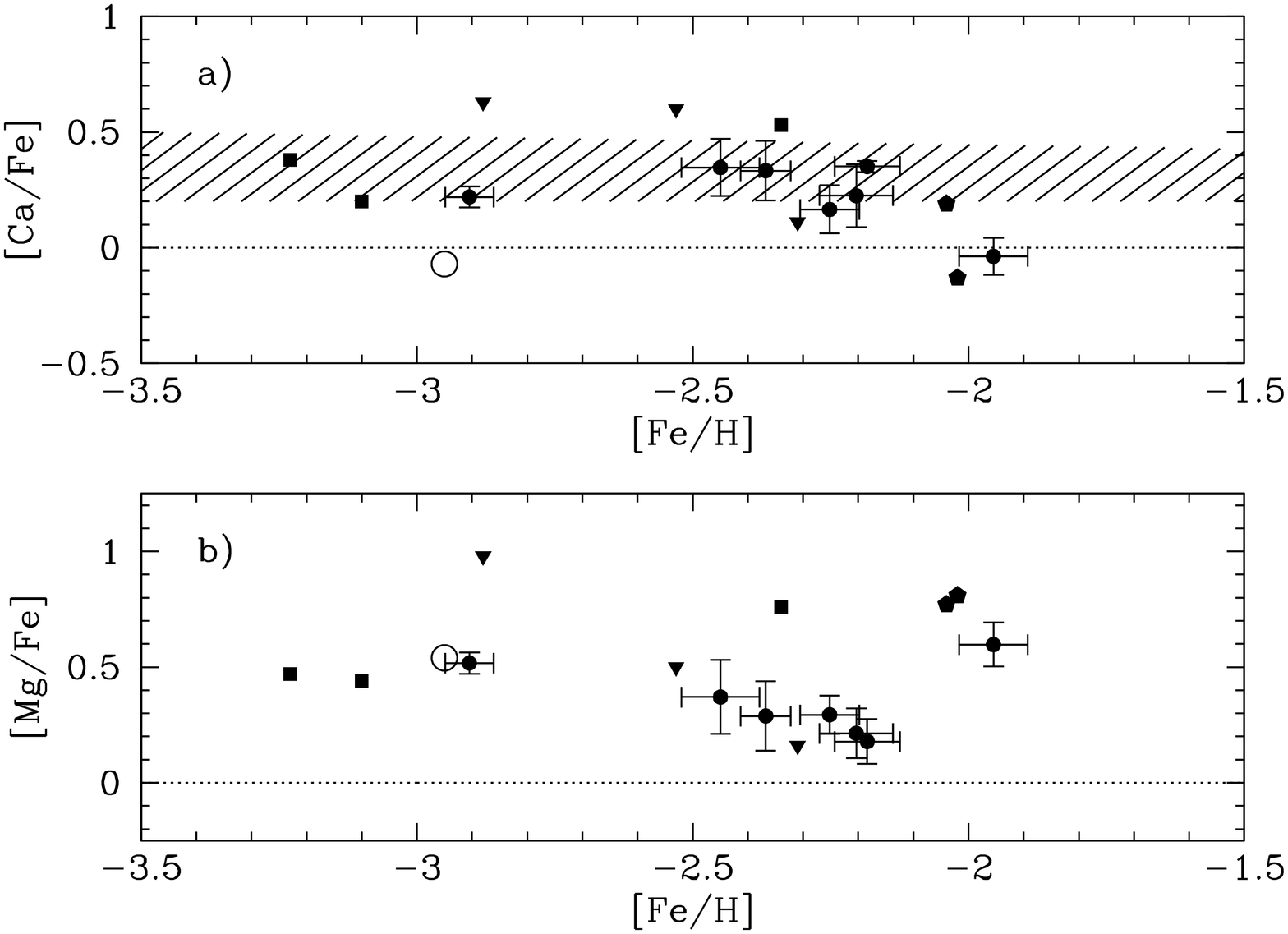}}
\caption{{\bf a)} A comparison of [Ca/Fe] vs, [Fe/H] for
  stars in several ultra-faint dSph galaxies. Bo\"otes\,I ($\bullet$
  with error-bars) Dra-119 \citep[$\bigcirc$][]{fulbright2004Draco},
  Hercules \citep[filled pentagons,][]{koch2008Her}, Ursa Major\,II
  (filled squares) and Coma Berenices (filled, inverted triangles)
  \citep[both from][]{frebel2009CBU}.  The hashed area shows the
  general trend for halo stars in the Milky Way. {\bf b)} The same
  stars as in a) but for [Mg/Fe] vs [Fe/H].}
          \label{camgfe.fig}
   \end{figure}

\citet{koch2008Her} have observed stars in the Hercules dSph galaxy
and found atypical abundance ratios. The Hercules dSph galaxy is one
of the ultra-faint, newly discovered dSph galaxies
\citep[][]{belokurovcats,aden}. One star in Draco, a classical dSph
galaxy, also shows this unusual abundance pattern
\citep[Dra-119][]{fulbright2004Draco}. We find one star in the
Bo\"otes\,I dSph galaxy that clearly shows the same unusual [Mg/Ca]
pattern, Boo-127 (Fig.\,\ref{mgca.fig}). It is also possible that
Boo-094 shows similar traits but not as clearly.

It is clear that Her-2 and Her-3, Dra-119, and Boo-127 all stand out
very clearly from the common trend. Even with large errors in the
stellar parameters, Boo-127 cannot be made compatible with the general
trend found for the other RGB stars in the Bo\"otes\,I dSph
galaxy. Note also that we have 3-4 Mg lines in our abundance analysis,
while \cite{koch2008Her} only used one. Our four lines show consistent
and high Mg abundances (cf. Table\,1). Thus our results confirm and
strengthen the results found for the Draco and Hercules dSph galaxies.

Additionally, \citet{koch2008Her} and \citet{fulbright2004Draco} find
that Her-2, Her-3, and Dra-119, respectively, have extremely weak or
nonexistent Ba\,{\sc ii} lines resulting in upper limits on the
[Ba/H] abundances. For the stars in the Bo\"otes\,I dSph galaxy, this
does not appear to be the case (compare Fig.\,\ref{fig:spectra}). We
have not derived Ba\,{\sc ii} abundances for Boo-094, because on closer
inspection, the lines are not free from blemishes, but they are clearly
visible in the stellar spectra.

However, for metal-poor stars, like Boo-094, the S/N of the spectrum
is clearly important for a positive detection of the Ba\,{\sc ii} line
at 649.689\,nm (compare Fig.\ref{fig:spectra}).  A comparison with
Fig.\,2 in \citet{koch2008Her} shows that it might be possible that
the line is buried in the noise. Those spectra also have somewhat
lower resolution.  Koch (2009, priv.com.) confirms that this might be
a possibility, but that it is unlikely that, if the line was present
at a standard level, it could be completely veiled by the noise and
lower resolution; hence, the Ba abundance would still be low.  Higher
resolution, higher S/N spectra of the stars in the Hercules dSph are
needed to fully settle the issue.

In supernova explosions, freshly synthesized elements are ejected into
the nearby interstellar medium. The yields of different elements
depend on the mass of the star that is the progenitor
\citep[e.g.][]{woosley1995,1999ApJS..125..439I}. In models of galactic
chemical evolution, it is often assumed that the ejecta from the
supernova become well-mixed quickly. This is the instantaneous
recycling approximation \citep[e.g.][]{pagel1997}. This works well
when we explore the later phases of galactic chemical evolution or
study galaxies as a whole; however, in situations where only one or
few supernova have had the chance to enrich the interstellar medium,
the gas will not be well-mixed, and we might therefore see an atypical
composition of elemental abundances in a single star
\citep[][]{karlsson2005}. Few studies have been done of the effect on
elemental abundances in low mass systems such as the ultra-faint dSph
galaxies. As opposed to the more robust predictions from models of
galactic chemical evolution that concerns giant galaxies such as the
Milky Way, models for dSph galaxies will be highly vulnerable to any
uncertainties in the supernova yields used for the modelling. As such
yields remain uncertain, we can only at this point speculate on the
origin of the atypical abundance ratios observed.

\citet{koch2008Her} speculate that, if the enrichment histories of the
ultra-faint dSph galaxies are largely dominated by inhomogeneous
evolution, considerable star-to-star scatter should be observed in
these systems. Apart from Boo-127 (and possibly Boo-094), the remaining
stars in Bo\"otes\,I show considerable homogeneity in their derived Mg
and Ca abundances (Fig.\,\ref{camgfe.fig}). Also for the two faint
systems Coma Berenices and Ursa Major\,II, \citet{frebel2009CBU} find
very homogeneous [Mg/Ca] abundance ratios (Fig.\,\ref{mgca.fig}) and
fairly homogeneous abundance trends for the $\alpha$-elements
(Fig.\,\ref{camgfe.fig}). These findings seem to suggest that the
ultra-faint dSph galaxies are not{\bf , after all,} different from the
classical dSph galaxies. Indeed, Draco, a classical dSph, has one odd
star \citep{fulbright2004Draco}, and the remainder are normal
\citep[Compare, e.g., the recent results in][which shows that eight of
the bright red giants in Draco all have normal Mg to Ca ratios.]{2009ApJ...701.1053C}.

\begin{acknowledgements}

       S.F. is a Royal Swedish Academy of Sciences Research Fellow
       supported by a grant from the Knut and Alice Wallenberg
       Foundation. K.E. gratefully acknowledges support from the
       Swedish Research Council.  M.I.W. is supported by a Royal
       Society University Research Fellowship.

The authors wish to recognize and acknowledge the very significant
cultural role and reverence that the summit of Mauna Kea has always
had within the indigenous Hawaiian community.  We are most fortunate
to have the opportunity to conduct observations from this mountain.

This publication makes use of data products from the Two Micron All
Sky Survey, which is a joint project of the University of
Massachusetts and the Infrared Processing and Analysis
Center/California Institute of Technology, funded by the National
Aeronautics and Space Administration and the National Science
Foundation.

\end{acknowledgements}

\bibliographystyle{aa}
\bibliography{referenser}

\end{document}